\documentclass[a4paper]{IEEEtran}

\usepackage{graphicx,url}
\usepackage{listings}
\usepackage{xcolor}
\usepackage{amssymb}
\usepackage{color}
\usepackage[british]{babel}   

\newcommand{\comment}[1]{}

\begin{document}
\title{Bounded Model Checking of\\C$++$ Programs Based on the Qt Framework  (extended version)}

\author{
\IEEEauthorblockN{Felipe R. M. Sousa, Lucas C. Cordeiro, and Eddie B. de Lima Filho\\}
\IEEEauthorblockA{Federal University of Amazonas - UFAM \\
									Manaus-AM, Brazil 69077--000\\
                  Email: \{felipemonteiro,lucascordeiro,eddie\}@ufam.edu.br}
}

\maketitle

\begin{abstract}
The software development process for embedded systems is getting faster and faster, which generally incurs an increase in the associated complexity. As a consequence, consumer electronics companies usually invest a lot of resources in fast and automatic verification processes, in order to create robust systems and reduce product recall rates. Because of that, the present paper proposes a simplified version of the Qt framework, which is integrated into the Efficient SMT-Based Bounded Model Checking tool to verify actual applications that use the mentioned framework. The method proposed in this paper presents a success rate of 94.45\%, for the developed test suite.
\end{abstract}


\IEEEpeerreviewmaketitle

\section{Introduction}

Consumer electronics companies increasingly invest effort and time to develop fast and cheap alternatives for verifying correctness in their systems, in order to avoid financial losses~\cite{Berard10}. Among such alternatives, one of the most effective and less expensive way is the model checking~\cite{Clarke99} approach. However, despite its advantages, there are many systems that could not be automatically verified, due to the unavailability of verifiers that support certain types of languages and frameworks. For instance, the Java PathFinder is able to verify Java code, based on byte-code~\cite{NASA07}, but it does not support verification of Java applications, which rely on the Android operating system. Indeed, it is true unless an abstract representation of the standard libraries (operational model), which conservatively approximates their semantics, is available.

The present work identifies the main Qt features used in real applications and, based on that, creates an operational model, which provides a way to analyse and check properties related to those features. The developed algorithms were integrated into a checker that uses Bounded Model Checking (BMC) based on Satisfiability Modulo Theories (SMT), known as Efficient SMT-based Context-Bounded Model Checker (ESBMC$++$)~\cite{ECBS13}, in order to verify specific properties in ANSI-C/C$++$ programs.
Although the combination of ESBMC$++$ and operational models has been applied to verify C$++$ programs~\cite{ECBS13}, in this work, additional models are developed in order to identify Qt framework structures and to verify specific properties related to such structures, via pre- and post-conditions. Given the current knowledge in software verification, there is no other model checker that applies BMC techniques to verify programs based on the Qt framework, regarding consumer electronics devices.

\section{Efficient SMT-Based Bounded Model Checking} 
\label{sec:esbmc}

ESBMC$++$ is a Context-Bounded Model Checker based on SMT solvers, which is used for ANSI-C/C$++$ programs~\cite{ECBS13}. ESBMC$++$ verifies single- and multi-threaded programs and checks for properties related to arithmetic under- and overflow, division by zero, out-of-bounds index, pointer safety, deadlocks, and data races. In ESBMC$++$, the verification process is completely automated and does not require the user to annotate programs with pre- or post-conditions.

\begin{figure}[htb]
  \centering
  \includegraphics[width=3in]{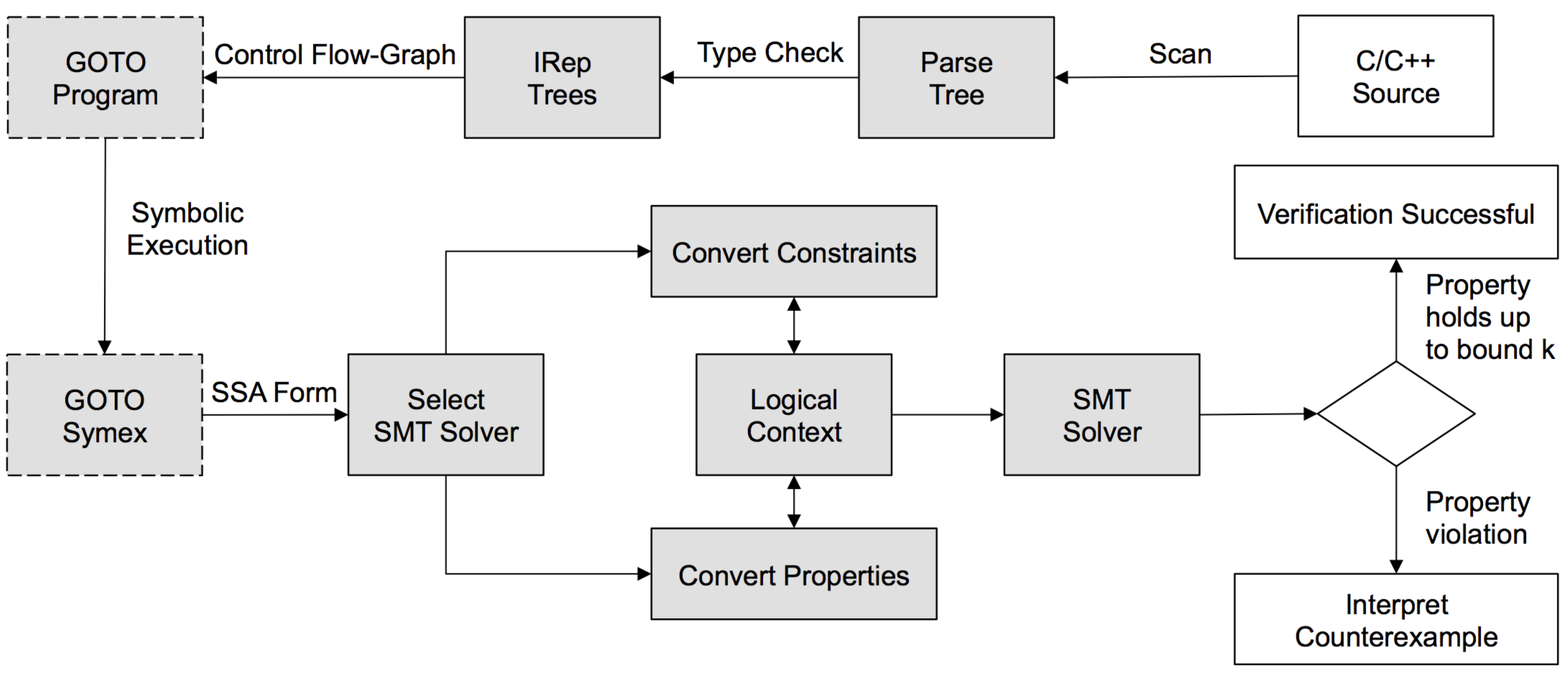}
  \caption{Overview of the ESBMC$++$ architecture.}
  \label{figure:Fig1}
\end{figure}

Fig.~\ref{figure:Fig1} shows the ESBMC$++$ architecture used for verifying C/C$++$ programs. ESBMC$++$ converts ANSI-C/C$++$ programs into equivalent \textit{GOTO-programs}, which simplify statement representations ({\it e.g.}, replacement of \textit{while} by \textit{if} and \textit{goto} statements). The \textit{GOTO-program} is symbolically executed by the \textit{GOTO-symex}, which generates a Single Static Assignment form that is later converted into an SMT formula and then checked by an SMT solver. If a property violation is found, a counterexample is provided by ESBMC$++$, which assigns values to the program variables to reproduce the error.

\section{Verifying C$++$ Programs Based on the Qt Cross-Platform Framework} 
\label{sec:ver}

In the first step of the verification process, the Qt program is converted into an intermediate representation (IRep) tree. However, since Qt is a robust framework, the set of standard Qt libraries contains hierarchical and complex structures, which make Qt programs verification an unfeasible task. Due to this particular reason, the use of an operational model, written in C$++$ and containing only structures needed for verifying properties related to the Qt framework, represents a feasible alternative. Additionally, the use of assertions is indispensable for verifying properties related to methods from the Qt framework and their execution, which is not covered by standard libraries. Such assertions are integrated into respective methods, in order to detect violations related to the incorrect use of the Qt framework. In summary, ESBMC$++$ is able to verify specific properties of the operational model. For example, the proposed methodology can check if a parameter, which represents a certain time period, is positive.

\begin{figure}[htb]
\centering
\includegraphics[width=2.5in]{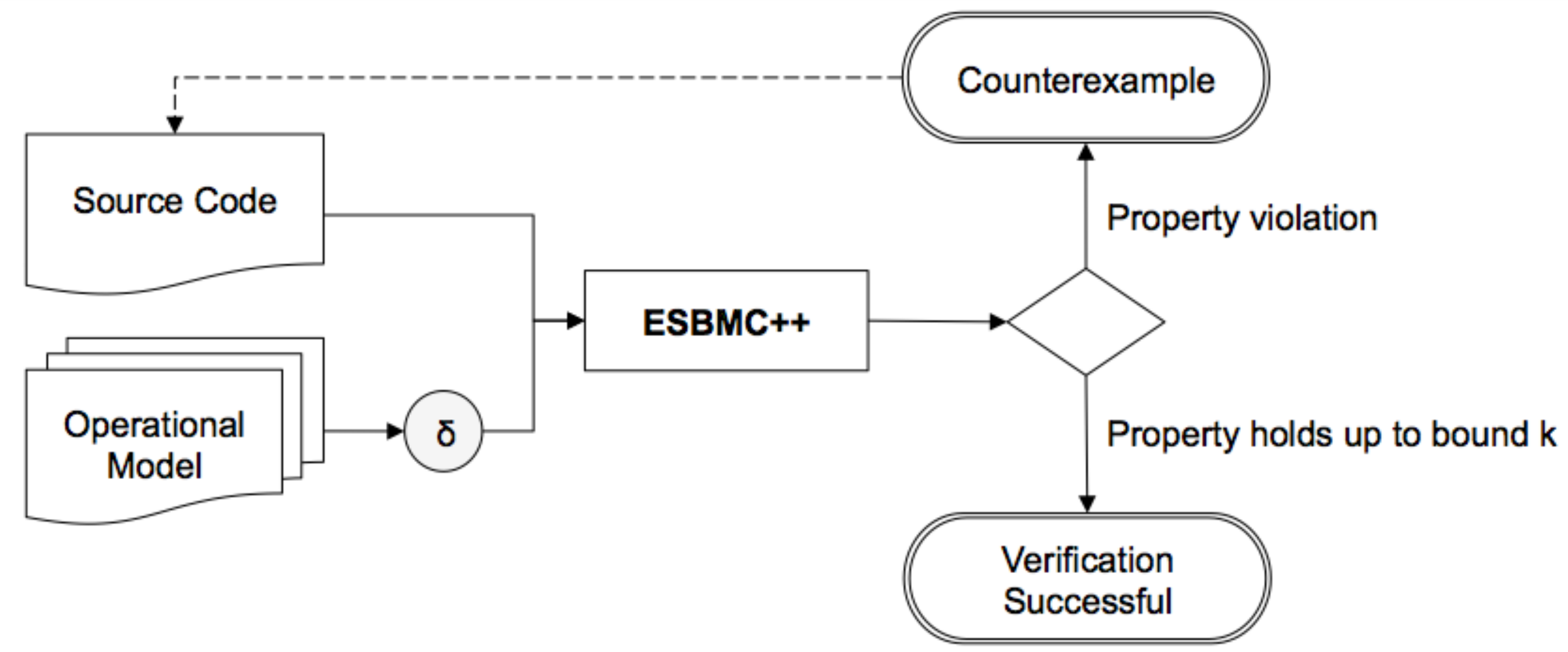}
\caption{Verification process for Qt programs using an operational model.}
\label{fig:Fig2}
\end{figure}

Fig.~\ref{fig:Fig2} shows the proposed verification process for Qt programs. In the first step, the model is connected to ESBMC$++$ and the Qt program to be verified, via a parameter $\delta$\footnote{{\tt-}I ~/libraries/Qt/}. After that, ESBMC$++$ automatically checks all pre- and post-conditions and, if there is no bug, it claims {\it Verification Successful} up to the analysed depth; otherwise, ESBMC$++$ returns a counterexample, reporting the line containing the error, the violated property, and the execution steps.

Based on the framework documentation~\cite{Qt15}, the operational model was developed, which considers the structure of each library and its classes, including attributes, method signatures, and function prototypes. From this simplified structure, assertions are integrated into the operational model for ensuring that each property is formally checked. Indeed, there are many properties to be verified, such as invalid memory access, negative time-period values, access to missing files, and null pointers. Additionally, there are pre- and post-conditions, which are necessary for the correct execution of QT methods.

\begin{figure}[htb]
\centering
\begin{minipage}{0.3\textwidth}
\begin{lstlisting}
	QList<int> mylist;
	mylist.push_front(300);
	assert(mylist.front() == 300);
	mylist.push_front(200);
	assert(mylist.front() == 200);
\end{lstlisting}
\end{minipage}
\caption{Code fragment using \textit{QList} class}
\label{fig:Fig3}
\end{figure}

In the example shown in Fig.~\ref{fig:Fig3}, the method $front$() handles a precondition~\cite{Qt15}. Indeed, an assertion was added to check whether the respective list is not empty (line $13$ of Fig.~\ref{fig:Fig4}) and contains the assigned value. When the $front$() method is called, ESBMC$++$ interprets its behavior as implemented in the operational model. As shown in Fig.~\ref{fig:Fig3}, the operation is valid and, consequently, the assertion evaluates to \textit{true}; however, if an invalid operation is performed, like an empty list calling $front$(), then the assertion would evaluate to \textit{false}. In that case, ESBMC$++$ would return a counterexample with all execution steps needed to reproduce the violation, in addition to the error described in the respective assertion.

\begin{figure}[htb]
\centering
\begin{minipage}{0.35\textwidth}
\begin{lstlisting}
template<class T>
class QList {
...
	void push_front( const value_type& x ){
		if(this->_size != 0) {
			for(int i = this->_size -1; i > -1; i--)
				this->_list[i+1] = this->_list[i];
		}
		this->_list[0] = x;
		this->_size++;
	}
	T& front() {
			__ESBMC_assert(!isEmpty(),
					``The list must not be empty'');
			return this->_list[0];
	}
...
}	
\end{lstlisting}
\end{minipage}
\caption{Operational model for the $push\_front$() and $front$() methods}
\label{fig:Fig4}
\end{figure}
Nonetheless, some methods not only contain properties that must be handled as preconditions, but also properties that are considered as postconditions. For instance, in the code fragment shown in Fig.~\ref{fig:Fig3}, an element is inserted into the beginning of a certain list ($mylist$) and, later on, the first element of the same list is checked with an assertion. From the operational model of $push\_front$(), in Fig.~\ref{fig:Fig4}, it is clear that if only preconditions are checked, then there is no evidence that elements are properly inserted into the respective list. This way, one needs to simulate the behavior of the respective method to consistently verify properties related to the manipulation or storage of values in a container. As a consequence, the operational model must strictly follow the specification described in the official documentation~\cite{Qt15}.


\section{Experimental Evaluation} 
\label{sec:dis}

All experiments were conducted on an otherwise idle Intel Core $i7$-$4790$, with $3.60$ GHz clock and $16$ GB of RAM, running Ubuntu OS ($64$ bits) and ESBMC$++$\footnote{The tool and benchmarks are available at http://www.esbmc.org} $1.25.4$.The time and memory limits, for each test case, were set to $600$ seconds and $16$ GB ($14$ GB of RAM and $2$ GB of virtual memory), respectively\footnote{{\tt--}unwind 10 {\tt--}no{\tt-}unwinding{\tt-}assertions {\tt-}I ~/libraries/Qt/ {\tt--}memlimit 14000000 {\tt--}timeout 600}. The indicated time periods were measured using the $time$ command. As an operational model was developed, benchmarks were included into an automatic test suite called \textit{esbmc-qt}, in order to validate this implementation. Currently, \textit{esbmc-qt} contains $52$ benchmarks ($1767$ code lines), which take about $48$ seconds to be verified. ESBMC$++$ presents a successful rate of $94.45$\% for the developed test suite, a ``false incorrect'' rate of $1.85$\%, which occurs when there is no error and ESBMC$++$ finds a violation, and a ``failed'' rate of $3.70$\%, which happens when ESBMC$++$ crashes during verification.

\section{Conclusion}
\label{conclusao}
%
This paper proposes an approach to verify C$++$/Qt programs using an operational model, which includes pre- and post-conditions, simulation features ({\it e.g.}, how element values of containers are manipulated and stored), and also how those are used in order to verify Qt applications, in consumer electronics devices. The experimental results show the efficiency and effectiveness of this approach for verifying Qt programs and present, for the developed test suite, a success rate of $94.45$\%. As future work, more classes and libraries will be integrated into the developed operational model, in order to increase Qt framework coverage and validate its properties.

\small{\noindent \textbf{Acknowledgements.} Part of the results presented in this paper were sponsored by Samsung Eletr{\^o}nica da Amaz{\^o}nia Ltda. under the terms of Brazilian federal law No. 8.387/91 (SUFRAMA).}

\bibliographystyle{IEEEtran}
\bibliography{references}

\begin{thebibliography}{1}
\providecommand{\url}[1]{#1}
\csname url@samestyle\endcsname
\providecommand{\newblock}{\relax}
\providecommand{\bibinfo}[2]{#2}
\providecommand{\BIBentrySTDinterwordspacing}{\spaceskip=0pt\relax}
\providecommand{\BIBentryALTinterwordstretchfactor}{4}
\providecommand{\BIBentryALTinterwordspacing}{\spaceskip=\fontdimen2\font plus
\BIBentryALTinterwordstretchfactor\fontdimen3\font minus
  \fontdimen4\font\relax}
\providecommand{\BIBforeignlanguage}[2]{{%
\expandafter\ifx\csname l@#1\endcsname\relax
\typeout{** WARNING: IEEEtran.bst: No hyphenation pattern has been}%
\typeout{** loaded for the language `#1'. Using the pattern for}%
\typeout{** the default language instead.}%
\else
\language=\csname l@#1\endcsname
\fi
#2}}
\providecommand{\BIBdecl}{\relax}
\BIBdecl

\bibitem{Berard10}
B.~Berard, M.~Bidoit, and A.~Finkel, \emph{Systems and Software Verification:
  Model-Checking Techniques and Tool}.\hskip 1em plus 0.5em minus 0.4em\relax
  Springer Publishing, 2010.

\bibitem{Clarke99}
E.~M. Clarke and et~al., \emph{Model Checking}.\hskip 1em plus 0.5em minus
  0.4em\relax Springer Publishing, 1999.

\bibitem{NASA07}
P.~C. Mehlitz, N.~Rungta, and W.~Visser, ``A hands-on java pathfinder
  tutorial,'' in \emph{{ICSE}}, 2013, pp. 1493--1495.

\bibitem{ECBS13}
M.~Ramalho, M.~Freitas, F.~Sousa, H.~Marques, L.~C. Cordeiro, and B.~Fischer,
  ``{SMT}-based bounded model checking of {C++} programs,'' in \emph{ECBS},
  2013, pp. 147--156.

\bibitem{Qt15}
{The Qt Framework}, http://www.qt.io/qt-framework/, {April}, 2015.

\end{thebibliography}

\end{document}